\global\def\draftcontrol{0}

   \def\versionno{ superpotential }

\catcode`\@=11

\newcommand\makepapertitle{\par

  \begingroup
    \renewcommand\thefootnote{\@fnsymbol\c@footnote}%
 \newpage
     \global\@topnum\z@   
     \@makepapertitle
     \thispagestyle{empty}\@thanks
  \endgroup
  \setcounter{footnote}{0}%
  \global\let\thanks\relax
  \global\let\makepapertitle\relax
  \global\let\@makepapertitle\relax
  \global\let\@thanks\@empty
  \global\let\@author\@empty
  \global\let\@date\@empty
  \global\let\@title\@empty
  \global\let\title\relax
  \global\let\author\relax
  \global\let\date\relax
  \global\let\and\relax
  \def\version{\let\version\@version\@gobble}
}
\def\@makepapertitle{%
  \newpage
   \ifnum\draftcontrol=1 {}
   \version\versionno
   \vskip 5.5em%
   \else
   \hfill\hbox to 3.5cm {\parbox{5cm}{\@pubnum}\hss}%
   \vskip 6.5em%
   \fi
   \begin{center}%
   \let \footnote \thanks
      {\hskip -0\textwidth \hbox to 1\textwidth%
        {\centerline{\Large\bf{\noindent%
    \parbox[t]{1.3\textwidth}{\begin{center}\@title\end{center}}}}}}%
     \vskip 1.5em%
     {\normalsize
       \lineskip .5em%
       \begin{tabular}[t]{c}%
         \@author
       \end{tabular}\par}%
     \vskip 1.5em%
     {\@bstract}%
     \end{center}%
     \vfill
     \@date%
     \vskip 1.5em%
   \par
}

\gdef\@pubnum{}
\def\pubnum#1{%
  \gdef\@pubnum{#1}}

\gdef\@bstract{}
\def\Abstract#1{%
  \gdef\@bstract{%
   \parbox{\textwidth-0pc}{%
   \centerline{\bf Abstract}\penalty1000
   \noindent
   \renewcommand\baselinestretch{1.0}
   {#1}}}
}

\gdef\@email{}
\def\email#1{%
   \gdef\@email{%
   Email: {\tt #1}}
}

\def\ps@paper{\let\@mkboth\@gobbletwo%
     \ifnum\draftcontrol=1
        \def\@oddfoot{\hbox to \textwidth{\tiny \versionno \hfil\tiny\draftdate}%
        \hskip -\textwidth \hbox to \textwidth{\hfil\rm\thepage\hfil}}%
     \else\def\@oddfoot{\hbox to \textwidth{\hfil\rm\thepage\hfil}}
     \fi
     \let\@evenfoot\@oddfoot
}

\def\body{\clearpage
          \pagestyle{paper}
        }

\def\@version#1{\ifnum\draftcontrol=1
\typeout{}\typeout{#1}\typeout{}
\vskip3mm\centerline{\hbox{\fbox{\normalsize{\tt DRAFT -- #1 -- }
                   {\draftdate}}}}\vskip3mm
\fi}
\let\version\@version
\long\def\eqlabel#1{\ifnum\draftcontrol=1
                    \tag@false  
                    \tag*{(\theequation) \hbox to -0.2cm{\hspace{0cm}\small{#1}\hss}}
                    \refstepcounter{equation}
                    \edef\@currentlabel{\theequation}
                    \ltx@label{#1}
                    \else
                    \label{#1}
                    \fi
                    }
\let\st@bibitem\@bibitem
\let\st@lbibitem\@lbibitem
\ifnum\draftcontrol=1
  \def\@bibitem#1{%
    \st@bibitem{#1}\a@@label{#1}\ignorespaces}
  \def\@lbibitem[#1]#2{%
    \st@lbibitem[#1]{#2}\a@@label{#2}\ignorespaces}
  \def\a@@label#1{%
    \gdef\a@lab{\smash{\normalfont\small#1}}
    \ifvmode
      \if@inlabel
        \global\setbox\@labels\hbox{%
          \llap{\a@lab\let\a@lab\relax
                \kern\@totalleftmargin\kern\marginparsep}%
          \box\@labels}%
      \fi
    \fi}
\fi

\documentclass[12pt,letterpaper]{article}
\usepackage{amsmath}
\usepackage{amsmath}
\usepackage{amsmath}
\usepackage{amssymb}
\usepackage{amssymb}
\usepackage{amssymb}
\usepackage{amsmath,bm,amsfonts,amssymb,array,calc,amsthm,rotating,cite}
\usepackage{epsfig,psfrag}
\usepackage{graphicx}
\usepackage{color}
\usepackage[colorlinks=true]{hyperref}
\usepackage[all]{xy}

\renewcommand\baselinestretch{1.25}
\renewcommand\section{\@startsection {section}{1}{\z@}%
                                   {-3.5ex \@plus -1ex \@minus -.2ex}%
                                   {2.3ex \@plus.2ex}%
                                   {\normalfont\large\bfseries}}
\renewcommand\subsection{\@startsection{subsection}{2}{\z@}%
                                   {-3.25ex\@plus -1ex \@minus -.2ex}%
                                   {1.5ex \@plus .2ex}%
                                   {\normalfont\normalsize\bfseries}}
\renewcommand\subsubsection{\@startsection{subsubsection}{3}{\z@}%
                                   {-3.25ex\@plus -1ex \@minus -.2ex}%
                                   {1.5ex \@plus .2ex}%
                                   {\normalfont\normalsize\it}}
\renewcommand\paragraph{\@startsection{paragraph}{4}{\z@}%
                                   {-3.25ex\@plus -1ex \@minus -.2ex}%
                                   {1.5ex \@plus .2ex}%
                                   {\normalfont\normalsize\bf}}
\renewcommand\subparagraph{\@startsection{subparagraph}{5}{\z@}%
                                   {-1.25ex\@plus -1ex \@minus -.2ex}%
                                   {0ex \@plus .2ex}%
                                   {\normalfont\normalsize\it}}

\numberwithin{equation}{section}

\renewcommand*\l@section[2]{%
  \ifnum \c@tocdepth >\z@
    \addpenalty\@secpenalty
    \addvspace{.5em \@plus\p@}%
    \setlength\@tempdima{1.5em}%
    \begingroup
      \parindent \z@ \rightskip \@pnumwidth
      \parfillskip -\@pnumwidth
      \leavevmode \bfseries
      \advance\leftskip\@tempdima
      \hskip -\leftskip
      #1\nobreak\hfil \nobreak\hb@xt@\@pnumwidth{\hss #2}\par
    \endgroup
  \fi}
\renewcommand*\l@subsection{\addvspace{.0em \@plus\p@}\@dottedtocline{2}{1.5em}{2.3em}}
\renewcommand*\l@subsubsection{\addvspace{-.2em \@plus\p@}\@dottedtocline{3}{3.8em}{3.2em}}

\definecolor{refcol}{rgb}{0.0,0.0,0.2}
\definecolor{eqcol}{rgb}{.2,0,0}
\definecolor{purple}{cmyk}{0,1,0,0}

\gdef\@citecolor{refcol} \gdef\@linkcolor{eqcol}
\gdef\@urlcolor{refcol}
\def\colorlinkspurple{\gdef\@urlcolor{purple}}
\def\colorlinksblue{\gdef\@urlcolor{blue}}
\def\colorlinksred{\gdef\@urlcolor{red}}

\def\revise#1       {\raisebox{-0em}{\rule{3pt}{1em}}%
                     \marginpar{\raisebox{.5em}{\vrule width3pt\
                     \vrule width0pt height 0pt depth0.5em
                     \hbox to 0cm{\hspace{0cm}{%
                     \parbox[t]{4em}{\raggedright\footnotesize{#1}}}\hss}}}}

\def\Omthree#1#2#3{{\Omega^{#1,#2}_{{#3}}}}

\def\sst{\scriptstyle}
\def\delh{{\delta_{\hat z}}}

\catcode`\@=12
\newcommand{\bqa}{\begin{eqnarray}}
\newcommand{\eqa}{\end{eqnarray}}
\begin{document}

\title{
Type II/F-theory Superpotentials and Ooguri-Vafa Invariants of
Compact Calabi-Yau Threefolds with Three Deformations}

\author{
Feng-Jun Xu,~~~Fu-Zhong Yang\footnote{Corresponding author~~~
E-mail: fzyang@gucas.ac.cn} \\[0.2cm]
\it College of Physical Sciences, Graduate University of Chinese Academy of Sciences\\
\it   YuQuan Road 19A, Beijing 100049, China}

\Abstract{~~~~We calculate the D-brane superpotentials for two
Calabi-Yau manifolds with three deformations by the generalized
hypergeometric GKZ systems, which give rise to the flux
superpotentials $\mathcal{W}_{GVW}$ of the dual F-theory
compactification on the relevant Calabi-Yau fourfolds in the weak
decoupling limit. We also compute the Ooguri-Vafa invariants from
A-model expansion with mirror symmetry, which are related to the
open Gromov-Witten invariants.}

\makepapertitle

\body

\version\versionno

\vskip 1em

\newpage

\section{Introduction}

~~~~Mirror symmetry has obtained much success in the $N=2$
supersymmetric theories. The exact non-perturbative holomorphic
prepotential\cite{Candelas,Bershadsky:1993cx} in such theories can
be obtained by topological string theory via mirror symmetry.

    Appearance of D-brane breaks the supersymmetry to
    $N=1$. When a D6-brane wrapped on a special lagrangian 3-cycle, the $N=1,~d=4$ superpotential term can be computed by the open topological string amplitudes
$\mathcal{F}_{g,h}$
 of the A-model as follows
 \bqa
 \eqlabel{top}
h\int
d^4xd^2\theta\mathcal{F}_{g,h}(\mathcal{G}^2)^g(\mathcal{F}^2)^{h-1}
 \eqa
where $\mathcal{G}$ is the gravitational chiral superfield and
$\mathcal{F}$ is the gauge chiral superfield. The formula
\eqref{top} at $g=0,~h=1$ leads to in F-terms of $N=1$ supersymetric
theories:
$$\int d^4xd^2\theta \mathcal{W}(\Phi)$$ Hence we can calculate these
superpotential by topological open-string theory.

  For non-compact Calabi-Yau manifolds, the refs.\cite{Aganagic:2000gs,Kachru:2000an,Kachru:2000ih,Aganagic:2001nx,Lerche:2002yw,Lerche:2002ck,Aganagic:2003db} studied the open-closed mirror
symmetry and its applications. In particular, the work
\cite{Aganagic:2000gs} constructed the classical A-brane geometry
with special Lagrangian submanifold and the work
\cite{Lerche:2002yw,Lerche:2002ck} introduced $N=1$ spacial geometry
and variation of mixed Hodge structure to calculate superpotentials.
However, computing such superpotentials for compact Calabi-Yau
threefold is a hard work. Recently, a progress on compact manifolds
came from
\cite{Walcher:2006rs,Morrison:2007bm,Krefl:2008sj,Knapp:2008uw},
which studied a class of involution branes independent of open
deformation moduli. Furthermore, there appeared some related works
on superpotential for compact Calabi-Yau manifolds depending on
open-closed deformation moduli
\cite{Alim:2009rf,Alim:2009bx,Alim:2010za,Li:2009dz,Jockers:2008pe,Jockers:2009mn,Jockers:2009ti,Alim:2011rp,Walcher:2009uj,Grimm:2008dq,Grimm:2009sy,Grimm:2010gk,Klevers:2011xs,Grimm:2009ef,Baumgartl:2007an,Baumgartl:2008qp,Baumgartl:2010ad,Shimizu:2010us,Fuji:2010uq}.
where the works \cite{Baumgartl:2007an,Baumgartl:2008qp} studied it by the conformal field theory and matrix factorization and the others considered with Hodge theoretic method.

    Open mirror symmetry also has important application in enumerative
geometry. The superpotentials of A-model at large radius region
count the disk invariants \cite{Aganagic:2000gs} which are related
to open Gromov-Witten invariants \cite{fukaya,L,KL,Jinzenji,Fang}.

   In this paper, we calculate D-brane superpotentials for compact Calabi-Yau threefolds with three deformation parameters by open-closed mirror symmetry and generalized GKZ system
   \cite{Batyrev:1993wa,Batyrev:1994hm,Hosono:1993qy,Hosono:1995bm,Jockers:2008pe,Lerche:2002ck,Lerche:2002yw}, which is closed related to variations of mixed
   Hodge structure on relative cohomology group. The compact
   Calabi-Yau threefolds we consider are constructed as Calabi-Yau
   hypersurfaces in ambient toric varieties. When including
   D-branes, the polyhedron associated to the above toric varieties can
   be extended to the so-called enhanced polyhedron with one dimension higher which gives rise
   to another toric variety and Calabi-Yau fourfold. There exists a
   duality between the type II compactification with brane on the
   threefold and the M/F-theory compactification on the Calabi-Yau
   fourfold without any branes but
   with fluxes
   \cite{Alim:2009bx,Alim:2009rf,Alim:2010za,Mayr:2001xk,Grimm:2009ef,Klevers:2011xs}.
    In the weak decoupling
   limit $g_s\rightarrow0$, the Gukov-Vafa-Witten superpotentials \cite{Gukov:1999ya} $\mathcal{W}_{GVW}$ of
   F-theory compactification on this fourfold agrees with superpotentials $\mathcal{W}$ of Type II
   compactification threefold with branes at lowest order in $g_s$
   \cite{Alim:2009bx,Alim:2010za,Jockers:2009ti,Berglund:2005dm,Grimm:2009ef,Klevers:2011xs}
   \bqa
\mathcal{W}_{GVW}=\mathcal{W}+\mathcal{O}(g_s)+\mathcal{O}(e^{-1/g_s})
   \eqa
Hence, in this limit, we can obtain the flux superpotential
$\mathcal{W}_{GVW}$ from the superpotential $\mathcal{W}$ which will
be given in this paper.

   In Sect. 2 we give a
   overview of $N=1$ special geometry and generalized GKZ system.
   In sect. 3 and sect. 4, we calculate the type II/F superpotentials for two compact Calabi-Yau threefold with three deformation parameters---$X_{24}(1,1,2,8,12)$ and
   $X_{12}(1,1,1,3,6)$, respectively.
   In these manifolds, We consider superpotential, mirror symmetry and Ooguri-Vafa invariants for D-brane with a single open deformation moduli. Sect. 5 is for
   summary.

\section{N=1 Special Geometry, Generalized GKZ System and Type II/F Superpotentials}

\subsection{ $N=1$ Special Geometry, Relative Periods and Type II/F Superpotentials}

 ~~~~Type II compactification theory is
described by an effective $N=1$ supergravity action with non-trivial
superpotentials on the deformation space $\mathcal{M}$ when adding
D-branes and background fluxes. For D6-brane wrapped the whole
Calabi-Yau threefold, the holomorphic Chern-Simons theory
\cite{Witten:1992fb} \bqa
  \mathcal{W}=\int_{X}\Omega^{3,0}\wedge \text{Tr}[A\wedge \bar{\partial}A+\frac{2}{3}A\wedge A\wedge A]
  \eqa
gives the brane superpotential $\mathcal{W}_{brane}$, where $A$ is
the gauge field with gauge group $U(N)$ for $N$ D6-branes. When
reduced dimensionally, the low dimenaional brane superpotentials can
be obtained as \cite{Lerche:2003hs,Aganagic:2000gs}
 \bqa
\mathcal{W}_{brane}
=N_{\nu}\int_{\Gamma^{\nu}}\Omega^{3,0}(z,\hat{z})=\sum_{\nu}N_{\nu}\Pi^{\nu}
 \eqa
 where $\Gamma^{\nu}$ is a special Lagrangian 3-chain and
$(z,\hat{z})$ are closed-string complex structure moduli and D-brane
moduli from open-string sector, respectively.

    The background fluxes $H^{(3)}=H_{RR}^{(3)}+\tau H_{NS}^{(3)}$, which take values in the integer cohomology group $H^3(X,\mathbb{Z})$, also break the supersymmetry $N=2$ to $N=1$.
 The $\tau=C^{(0)}+ie^{-\varphi}$ is the complexified Type IIB coupling field. Its
 contribution to superpotentials is \cite{Mayr:2000hh,Taylor:1999ii}
 \bqa
 \mathcal{W}_{flux}(z)\ =\int_{X}
 H_{RR}^{(3)}\wedge  \Omega^{3,0} = \sum_\alpha N_\alpha\cdot\Pi^\alpha(z)\ ,
 \qquad N_\alpha \in Z.
\eqa

  The contributions of D-brane and background flux (here the NS-flux ignored) give together the general
  form of superpotential as follow \cite{Lerche:2002ck,Lerche:2002yw}
  \bqa
\mathcal{W}(z,\hat{z})=\mathcal{W}_{brane}(z,\hat{z})+\mathcal{W}_{flux}(z)=\sum_{\gamma_{\Sigma}\in
H^3(Z^\ast,\mathcal{H})}N_{\Sigma}\Pi_{\Sigma}(z,\hat{z}) \eqa where
$N_{\Sigma}=n_{\Sigma}+\tau m_{\sigma}$, $\tau$ is the dilaton of
type II string and $\Pi_{\Sigma} $ is a relative periods defined in
a relative cycle $\Gamma\in H_3(X,D)$ whose boundary is wrapped by
D-branes and D is a holomorphic divisor of the Calabi-Yau space. In
fact, the two-cycles wrapped by the D-branes are holomorphic cycles
only if the moduli are at the critical points of the
superpotentials. Thus, the two-cycles are generically not
holomorphic. However, according to the arguments of
\cite{Lerche:2002ck,Lerche:2002yw,Jockers:2008pe}, the
non-holomorphic two-cycles can be replaced by a holomorphic divisor
D of the ambient Calabi-Yau space with the divisor D encompassing
the two-cycles.

Geometrically speaking, when varying the complex structure of
Calabi-Yau space, a generic holomorphic curve will not be
holomorphic with the respect to the new complex structure, and
becomes obstructed to the deformation of the bulk moduli. The
requirement for the holomorphy gives rise to a relation between the
closed and open string moduli. Physically speaking, it turns out
that the obstruction generates a superpotential for the effective
theory depending on the closed and open string moduli.

   The off-shell
tension of D-branes, $\mathcal{T}(z,\hat{z})$, is equal to the
relative period \cite{Witten:1997ep,Lerche:2002ck,Lerche:2002yw} \bqa
\Pi_{\Sigma}=\int_{\Gamma_{\Sigma}}\Omega(z,\hat{z}) \eqa
 which
measures the difference between the value of on-shell
superpotentials for the two D-brane configurations\bqa
\mathcal{T}(z,\hat{z})=\mathcal{W}(C^+)-\mathcal{W}(C^-) \eqa  with
$\partial \Gamma_{\Sigma}=C^+-C^-$. The  on-shell domain wall
tension is \cite{Alim:2010za} \bqa
T(z)=\mathcal{T}(z,\hat{z})\mid_{\hat{z}=\text{critic points}}\eqa
where the critic points correspond to $\frac{dW}{d\hat{z}}=0$
\cite{Witten:1997ep} and the $C^{\pm}$ is the holomorphic curves at
those critical points. The critical points are alternatively defined
as the Nother-Lefshetz locus \cite{Clemens}
 \bqa
\mathcal{N}=\{(z,\hat{z})\mid\pi(z,\hat{z};\partial\Gamma(z,\hat{z}))\equiv
0\}
 \eqa
where \bqa \eqlabel{relative}
\pi(z,\hat{z};\partial\Gamma(z,\hat{z}))=\int_{\partial\Gamma}\omega_{\hat{a}}^{(2,0)}(z,\hat{z}),~~~\hat{a}=1,...,\text{dim}(H^{2,0}(D))
\eqa and $\omega_{\hat{a}}^{(2,0)}$ is an element of the cohomology
group $H^{(2,0)}(D)$. At those critical points, the domain wall
tensions are also known as normal function giving the Abel-Jacobi
invariants \cite{Clemens,
Alim:2010za,Li:2009dz,Morrison:2007bm,Griffiths}

  The relative periods $\Pi_{\Sigma}$ are related to the variations of mixed Hodge structure
  on the relative cohomology group $H^3(X,D)$. Geometrically, $H^3(X,D)$ can
be viewed as the fiber of a complex vector bundle over the
deformation space $\mathcal{M}$. The space $\mathcal{M}$, in
general, can be expressed \cite{Alim:2009bx,Alim:2010za}
  as fibration $\hat{\mathcal{M}}\rightarrow \mathcal{M}\rightarrow
  \mathcal{M}_{CS}$, where the base
  $\mathcal{M}_{CS}$ corresponds to the complex structure
  deformation space
 of the family of Calabi-Yau threefold $X$ and the fiber
$\hat{\mathcal{M}}$ is the deformation space of the family of
divisor $\mathcal{D}$ specifying the embedding
$i:D(z,\hat{z})\rightarrow X(z)$, which defines the space
$\Omega(X,D)$ of the relative differential form via the exact
sequence \bqa 0\rightarrow \Omega(X,D)\rightarrow
\Omega(X)\rightarrow \Omega(D)\rightarrow 0,\eqa so the relative
cohomology group can be reprensented as \bqa \eqlabel{minimal}
 H^3(X,D)\simeq
\text{ker} (H^3(X)\rightarrow H^3(D))\oplus
 \text{coker}(H^2(X)\rightarrow H^2(D)).
 \eqa

     The variation of mixed Hodge structure can be expressed as follow \cite{Lerche:2003hs}
$$
 \xymatrix{ (\Omthree30X,0)\ar[r]^{\sst
\delta_z}\ar[dr]^{\sst \delh}& (\Omthree21X,0)\ar[r]^{\sst
\delta_z}\ar[dr]^{\sst \delh}& (\Omthree12X,0)\ar[r]^{\sst
\delta_z}\ar[dr]^{\sst \delh}&
(\Omthree03X,0)\ar[dr]^{\sst \delta_z,\delh} \\
&(0,\Omthree20D)\ar[r]^{\sst \delta_z,\delh}&
(0,\Omthree11D)\ar[r]^{\sst \delta_z,\delh}&
(0,\Omthree02D)\ar[r]^{\sst \delta_z,\delh}& 0}
$$ which can lead to a system of differential equation for the periods in
open-closed mirror symmetry \cite{Lerche:2003hs}:
 \bqa
\nabla_z\cdot \Pi_{\Sigma} &=
\big(\partial_{z}-\mathcal{A}_z\big)\cdot \Pi_{\Sigma}(z,\hat z)\
=&\ 0\\ \nabla_{\hat z}\cdot \Pi_{\Sigma}\ &=\ \big(\partial_{\hat
z}- \mathcal{A}_{\hat z}\big)\cdot\Pi_{\Sigma}(z,\hat z)\ =&\ 0\
 \eqa  where $\mathcal{A}$ are Gauss-Manin connection, which
can be chosen as flat connection \cite{Lerche:2002yw,Lerche:2003hs}
\bqa \big[\nabla_{z_i},\nabla_{\hat z_j}\big] =
\big[\nabla_{z_i},\nabla_{z_j}\big]\ = \big[\nabla_{\hat
z_i},\nabla_{\hat z_j}\big] = 0.
 \eqa
 This approach provides a powerful framework to study
 relative periods and off-shell
superpotentials \cite{Lerche:2002yw,Alim:2009bx,Alim:2011rp}.

    In A-model interpretation, the superpotential expressed in term of flat coordinates $(t,\hat{t})$, which relates to complex
structure parameters $(z,\hat{z})$ in B-model through mirror map, is
the generating function of the Ooguri-Vafa invariants
\cite{Aganagic:2000gs,Ooguri:1999bv,Lerche:2002ck,Alim:2009rf} \bqa
\mathcal{W}(t,\hat{t})=\sum_{\overrightarrow{k},\overrightarrow{m}}G_{\overrightarrow{k},\overrightarrow{m}}q^{d\overrightarrow{k}}\hat{q}^{d\overrightarrow{m}}=\sum_{\overrightarrow{k},\overrightarrow{m}}\sum_{d}n_{\overrightarrow{k},\overrightarrow{m}}\frac{q^{d\overrightarrow{k}}\hat{q}^{d\overrightarrow{m}}}{k^2}
\eqa where $q=e^{2\pi it}$, $\hat{q}=e^{2\pi i\hat{t}}$ and
$n_{\overrightarrow{k},\overrightarrow{m}}$ are Ooguri-Vafa
invariants\cite{Ooguri:1999bv} counting disc instantons in relative
homology class $(\overrightarrow{m},\overrightarrow{k})$, where
$\overrightarrow{m}$ represents the elements of $H_1(D)$ and
$\overrightarrow{k}$ represents an element of $H_2(X)$.
$G_{\overrightarrow{k},\overrightarrow{m}}$ are open Gromov-Witten
invariants \cite{fukaya,L,KL,Jinzenji,Fang}. From string world-sheet
viewpoint, these terms in the superpotential represents the
contribution from instantons of sphere and disk.

\subsection{Generalized GKZ system and Differential Operators}

 ~~~~The generalized hypergeometric systems originated from \cite{GKZ} and have been
applied in mirror symmetry
\cite{Hosono:1993qy,Hosono:1995bm,Batyrev:1993wa,Batyrev:1994hm,Hosono}
. The notation is as follows: $(X^{\ast},X)$ is the mirror pair of
compact Calabi-Yau threefold defined as hypersurfaces in toric
ambient spaces $(W^{\ast},W)$, respectively. The generators $l^a$ of
Mori cone of the toric variety
\cite{Fujino,Fujino2,Scaramuzza,Renesse} give rise to the charge
vectors of the gauged linear sigma model (GLSM)\cite{Witten:1993yc}.
$\bigtriangleup$ is a real four dimensional reflexive polyhedron.
$W=P_{\Sigma(\bigtriangleup)}$ is the toric variety with fan
$\Sigma(\bigtriangleup)$ being the set of cones over the faces of
$\bigtriangleup^\ast$. $\bigtriangleup^\ast$ is the dual polyhedron
and $W^\ast$ is the toric variety obtained from
$\Sigma(\bigtriangleup^\ast)$. The enhanced polyhedron
$\underline{\triangle^{\ast}}$ constructed from polyhedron
$\bigtriangleup^\ast$ is associated to $X_4^{\ast}$ on which the
dual F-theory compactify. The three-fold X on B-model side is
defined by $p$ integral points of $\triangle^{\ast}$ as
   the zero locus of the polynomial $P$ in the toric ambient space
   \bqa
P=\sum_{i=0}^{p-1}a_i\prod_{k=0}^{4}X_k^{\nu_{i,k}^{\ast}}
   \eqa
where the $X_k$ are coordinates on an open torus
$(\mathbb{C}^{\ast})^4\in W$ and $a_i$ are complex parameters
related to the complex structure of X. In terms of homogeneous
coordinates $x_j$ on the toric ambient space, it can be rewritten as
\bqa  \eqlabel{constraint}
 P=\sum_{i=0}^{p-1}a_i\prod_{\nu\in
\triangle}x_j^{\langle\nu,\nu_i^{\ast}\rangle+1}. \eqa

  The open-string sector from D-branes can be described by the family
  of hypersurfaces $\mathcal{D}$, which is defined as intersections
$P=0=Q(\mathcal{D})$. In toric variety, the $Q(\mathcal{D})$ can be
defined as \cite{Alim:2009rf,Alim:2010za} \bqa
Q(\mathcal{D})=\sum_{i=p}^{p+p'-1}a_iX_k^{\nu_{i,k}^{\ast}}\eqa
where additional $p'$ vertices $v_i^{\ast}$ correspond with the
monomials in $Q(\mathcal{D})$.

  When considering the dual F-theory
compactify on Four-fold $X_4$, the relevant Enhanced polyhedron
consists of extended vertices
\begin{equation}
\eqlabel{four} \underline{\overline{\nu}_i^{\ast}}=\left\{
\begin{aligned}
(\nu_i^{\ast},0)~~i=0,...,p-1 \\
(\nu_i^{\ast},1)~i=p,...,p+p'-1.
\end{aligned}\right.
\end{equation}

   The period integrals can be written as
   \bqa
   \Pi_i=\int_{\gamma_i}\frac{1}{P(a,X)}{\prod_{j=1}^n \frac{d
   X_j}{X_j}}.
   \eqa
   According to the refs.\cite{Batyrev:1993wa,Batyrev:1994hm}, the
   period integrals can be annihilated by differential operators
   \bqa
   \eqlabel{operator}
   \begin{split}
   &\mathcal{L}(l)=\prod_{l_i>0}(\partial_{a_i})^{l_i}-\prod_{l_i<0}(\partial_{a_i})^{l_i}\\
   &\mathcal{Z}_k=\sum_{i=0}^{p-1}\nu_{i,k}^{\ast}\vartheta_i,
   \qquad \mathcal{Z}_0=\sum_{i=0}^{p-1}\vartheta_i-1
   \end{split}
   \eqa
where $\vartheta_i=a_i\partial_{a_i}$. As noted in
refs.\cite{Hosono:1993qy}, the equations $\mathcal{Z}_k\Pi(a_i)=0$
reflex the invariance under the torus action, defining torus
invariant algebraic coordinates $z_a$ on the moduli space of complex
structure of $X$: \bqa \eqlabel{coordinates}
z_a=(-1)^{l_0^{a}}\prod_{i}a_i^{l_i^{a}}\eqa where $l_a$,
~$a=1,...,h^{2,1}(X)$ is generators of the Mori cone, one can
rewrite the differential operators $\mathcal{L}(l)$ as
\cite{Batyrev:1994hm,Hosono:1993qy,Alim:2010za}
    \bqa
    \mathcal{L}(l)=\prod_{k=1}^{l_0}(\vartheta_0-k)\prod_{l_i>0}\prod_{k=0}^{l_i-1}(\vartheta_i-k)-(-1)^{l_0}z_a\prod_{k=1}^{-l_0}(\vartheta_0-k)\prod_{l_i<0}\prod_{k=0}^{l_i-1}(\vartheta_i-k).
    \eqa

   The solution to the GKZ system can be written as \cite{Batyrev:1994hm,Hosono:1993qy,Alim:2010za} \bqa
B_{l^a}(z^a;\rho)=\sum_{n_1,...,n_N\in Z_0^{+}}\frac{\Gamma(1-\sum_a
l_0^a(n_a+\rho_a))}{\prod_{i>0}\Gamma(1+\sum_a
l_i^a(n_a+\rho_a))}\prod_a z_a^{n_a+\rho_a}.\eqa

 In this paper we consider the family of divisors $\mathcal{D}$ with a single open
 deformation moduli $\hat{z}$
  \bqa
    x_1^{b_1}+\hat{z}x_2^{b_2}=0
  \eqa where $b_1,~b_2$ are some appropriate integers. However, in\cite{Grimm:2008dq,Grimm:2010gk}, they considered another approach which blows up along the curve C and replaces the pair(X,C) with a non-Calabi-Yau manifold $\widehat{X}$. The relative 3-form $\underline{\Omega}:=(\Omega_{X}^{3,0},0)$ and the relative
periods satisfy a set of differential equations
\cite{Lerche:2002ck,Lerche:2002yw,Alim:2009bx,Alim:2010za,Jockers:2008pe}
 \bqa
\mathcal{L}_a(\theta,\hat{\theta})\underline{\Omega}=d\underline{\omega}^{(2,0)}~\Rightarrow~\mathcal{L}_a(\theta,\hat{\theta})\mathcal{T}(z,\hat{z})=0.
\eqa with some corresponding two-form $\underline{\omega}^{(2,0)}$.
The differential operators $\mathcal{L}_a(\theta,\hat{\theta})$ can
be expressed as \cite{Alim:2010za} \bqa
\mathcal{L}_a(\theta,\hat{\theta}):=\mathcal{L}_a^{b}-\mathcal{L}_a^{bd}\hat{\theta}
\eqa for $\mathcal{L}_a^{b}$ acting only on bulk part from closed
sector, $\mathcal{L}_a^{bd}$ on boundary part from open-closed
sector and $\hat{\theta}=\hat{z}\partial_{\hat{z}}$. The explicit
form of these operators will be given in following model. From the
\eqref{relative} one can obtain \bqa \eqlabel{sub} 2\pi
i\hat{\theta}\mathcal{T}(z,\hat{z})=\pi(z,\hat{z}) \eqa for only the
family of divisors $\mathcal{D}$ depending on the $\hat{z}$. From
above one can obtain differential equation with the inhomogeneous
term $f_a(z)$ at the critical points \bqa
\mathcal{L}_a^{b}T(z)=f_a(z) \eqa and \bqa 2\pi i
f_a(z)=\mathcal{L}_a^{bd}\pi(z,\hat{z})|_{\hat{z}=\text{critic
points}}
 \eqa

\section{Superpotentials of Hypersurface $X_{24}(1,1,2,8,12)$}
 ~~~~The $X_{24}(1,1,2,8,12)$ is defined as a degree 24 hypersurface
 in the ambient toric variety $W=P_{\Sigma(\triangle)}$ with the
 vertices of the polyhedron $\triangle$
 \bqa
 \begin{split}
&\nu_1=(1,1,1,1),~~\nu_2=(-23,1,1,1),~~\nu_3=(1,-11,1,1),~~\nu_4=(1,1,-2,1)\\
&\nu_5=(1,1,1,-1)
\end{split}
\eqa
  The vertices of
dual polyhedra $\triangle^{\ast}$ are \bqa
\begin{split}
&\nu_1^{\ast}=(1,2,8,12),~~\nu_2^{\ast}=(-1,0,0,0),~~\nu_3^{\ast}=(0,-1,0,0),~~\nu_4^{\ast}=(0,0,-1,0)\\
&\nu_5^{\ast}=(0,0,0,-1),~~\nu_6^{\ast}=(0,1,4,6),~~\nu_7^{\ast}=(0,0,2,3).
\end{split}
\eqa When considering the four-fold on which the dual F-theory
compactification, the extend vertices of the enhanced polyhedron
$\underline{\triangle^{\ast}}\supset\triangle^{\ast}$ can be
constructed according to \eqref{four} with two extra points as
follows \bqa
\underline{\nu_i^{\ast}}=(\nu_i^{\ast};0),~\underline{\nu_8^{\ast}}=(\nu_1^{\ast};1),~\underline{\nu_9^{\ast}}=(\nu_2^{\ast};1),~i=0,...,7.
\eqa

    The toric hypersurface, according to \eqref{constraint},
is the zero locus of polynomial $P$\bqa
\begin{split}
P&=a_1x_1^{24}+a_2x_2^{24}+a_3x_3^{12}+a_4x_4^3+a_5x_5^{2}+a_0
x_1x_2x_3x_4x_5+a_6 x_1^{12}x_2^{12}+a_7 x_1^6x_2^6x_3^6 \\
&=x_1^{24}+x_2^{24}+x_3^{12}+x_4^{3}+x_5^{2}+\psi
x_1x_2x_3x_4x_5+\phi x_1^6x_2^6x_3^6+\chi x_1^{12}x_2^{12}
\end{split}\eqa
where the second equation is rescaled and expressed by
$\psi=z_1^{-\frac{1}{6}}z_2^{-\frac{1}{24}}z_3^{-\frac{1}{12}}$,
$\phi=z_2^{-\frac{1}{4}}z_3^{-\frac{1}{2}}$ and
$\chi=z_2^{-\frac{1}{2}}$ in terms of torus invariant algebraic
coordinates \eqref{coordinates}.
    The GLSM charge vectors $l_a$ are the generators of the Mori cone as follows \cite{Hosono:1993qy}
\begin{equation}
\begin{tabular}{c|c c c c c c c c}
~  & $0$ & $1$& $2$& $3 $ & $4$ & $5$ & $6$ & $7$ \\\hline
$l_1$ & $-6$ & $0$ & $0$& $0$ & $2$ & $3$ & $0$ & $1$ \\
$l_2$ & $0$ & $1$& $1$  & $0$& $0$ & $0$ & $-2$ & $0$ \\
$l_3$ & $0$ &$0$& $0$ & $1$ & $0$ & $0$ & $1$ & $-2$.
\end{tabular}
\end{equation}

    The mirror manifolds can be constructed as an orbifold by the
Greene-Plesser orbifold group acting as $x_i\rightarrow
\lambda_k^{g_{k,i}}x_i$ with weights \bqa \mathbb{Z}_6:
~g_1=(1,-1,0,0,0),~~~\mathbb{Z}_6:~g_2=(1,0,-1,0,0),~~~\mathbb{Z}_3:~g_3=(1,0,0,-1,0)\eqa
where we denotes $\lambda_{1,2}^6=1 ~\text{and} ~\lambda_{3}^3=1 $.
   We consider the following curves
  \bqa
 C_{\alpha,\pm}=\{x_2=\xi
  x_1,x_3=0,x_4^3=-x_5^2-\xi^{12}\chi x_1^{24}\}, ~\xi^{24}=-1
  \eqa
which are on the family of divisor  \bqa \eqlabel{Divisor1}
Q(\mathcal{D})=x_2^{24}+\hat{z}x_1^{24}\eqa at the critical points
$\hat{z}=1$. The surface defined by the intersection
$P=0=Q(\mathcal{D})$ is a $K3$ surface with the equation:\bqa
  P_{\mathcal{D}}=(x_2')^{12}+x_3^{12}+x_4^{3}+x_5^2+\psi'x_2'x_3x_4x_5+\phi'
  (x_2')^6x_3^6+\chi'(x_2')^{12}=0
  \eqa
where $x_2^{'}=x_2^2$,
$\psi^{'}=u_1^{-\frac{1}{6}}u_2^{-\frac{1}{24}}u_3^{-\frac{1}{12}}$,
$\phi^{'}=u_2^{-\frac{1}{4}}u_3^{-\frac{1}{2}}$ and
$\chi'=u_2^{-\frac{1}{2}}$ are expressed in terms of new parameters
as \bqa u_1=z_1 \qquad u_2=\frac{-z_2}{\hat{z}}(1-\hat{z})^2 \qquad
u_3=z_3\eqa

   The GLSM charge vectors for this K3 manifold are
\begin{equation}
\begin{tabular}{c|c c c c c c c}
~  & $0$ & $1$& $3 $ & $4$ & $5$ & $6$ & $7$ \\\hline
$\hat{l}_1$ & $-6$ & $0$ & $0$ & $2$ & $3$ & $0$ & $1$ \\
$\hat{l}_2$ & $0$ & $2$ & $0$& $0$ & $0$ & $-2$ & $0$ \\
$\hat{l}_3$ & $0$ & $0$ & $1$& $0$ & $0$ & $1$ & $-2$
\end{tabular}
\end{equation}
By the generalized GKZ system, the period on the K3 surface has the
form \bqa
\pi=\frac{c}{2}B_{\{\hat{l}_1,\hat{l}_2,\hat{l}_3\}}(u_1,u_2,u_3;\frac{1}{2},\frac{1}{2},0)=-\frac{4c}{\pi^{\frac{3}{2}}}\sqrt{u_1u_2}u_3+\mathcal{O}((u_1u_2)^{3/2})
\eqa which vanishes at the critical locus $u_2=0$. According to
\eqref{sub}, the off-shell superpotentials can be obtained by
integrating the $\pi$:
  \bqa
  \mathcal{T}_a^{\pm}(z_1,z_2,z_3)=\frac{1}{2\pi
  i}\int\pi(\hat{z})\frac{d\hat{z}}{\hat{z}}
\eqa with the appropriate integral constants\cite{Alim:2010za},
 the superpotentials can be chosen as  $\mathcal{W}^+=-\mathcal{W}^-$. In this convention, the on-shell
 superpotentials can be obtained as
\bqa 2\mathcal{W}^{+}=\frac{1}{2\pi
 i}\int_{-\hat{z}}^{\hat{z}}\pi(\zeta)\frac{d\zeta}{\zeta},~~~W^{\pm}(z_1,z_2,z_3)=\mathcal{W}^{\pm}(z_1,z_2,z_3)|_{\hat{z}=1}
\eqa
  Eventually, The superpotential are
\bqa
\begin{split}
&\mathcal{W}^{\pm}(z_1,z_2,z_3,\hat{z})=\sum_{n_1,n_2,n_3}\frac{\mp cz_1^{\frac{1}{2}+n_1}z_2^{\frac{1}{2}+n_2}z_3^{n_3}\hat{z}^{\frac{-1-2n_2}{2}}\Gamma(6n_1+4)}{\Gamma(2+2n_2)\Gamma(2+2n_1)\Gamma(\frac{5}{2}+3n_1)}\\
&\frac{\{(1-2n_2)
{}_2F_1(-\frac{1}{2}-n_2,-2n_2,\frac{1}{2}-n_2;\hat{z})+\hat{z}(1+2n_2){}_2F_1((\frac{1}{2}-n_2,-2n_2,\frac{3}{2}-n_2;\hat{z}))\}}{4\pi(-1+4n_2^2)\Gamma(1+n_3)\Gamma(n_3-2n_2)\Gamma(n_1-2n_3+\frac{3}{2})}
\end{split}
\eqa and can divide two parts \bqa
\mathcal{W}^{\pm}(z_1,z_2,z_3,\hat{z}')=W^{\pm}(z_1,z_2,z_3)+f(z_1,z_2,z_3,\hat{z}'),~~\hat{z}'=\hat{z}-1\eqa
where the $f(z_1,z_2,z_3,\hat{z}')$ are related to the open-string
parameter and vanish in the critical point, $W^{\pm}$ are the
on-shell superpotential as follows \bqa
W^{\pm}=\mp\frac{c}{8}B_{[l_1,l_2,l_3]}((z_1,z_2,z_3);\frac{1}{2},\frac{1}{2},0)\eqa
 substituting the vector $l_1,l_2,l_3$ in this hypersurface, the
 on-shell superpotentials are
 \bqa
 \begin{split}
W^{\pm}=&\frac{c}{8}\sum_{n_1,n_2,n_3}\frac{\mp z_1^{\frac{1}{2}+n_1}z_2^{\frac{1}{2}+n_2}z_3^{n_3}\Gamma(6(n_1+\frac{1}{2})+1)}{\Gamma(n_2+\frac{1}{2}+1)^2\Gamma(n_3+1)\Gamma(2(n_1+\frac{1}{2})+1)\Gamma(3(n_1+\frac{1}{2})+1)}\\
&\frac{1}{\Gamma(-2n_3+n_1+\frac{1}{2})+1)\Gamma(-2n_2+n_3)}.
 \end{split}
 \eqa

The additional GLSM charge vectors corresponding to the divisor
\eqref{Divisor1}
   are
       \begin{equation}
\begin{tabular}{c|c c c c c c c c c c}
 & $0$ & $1$& $2$& $3 $ & $4$ & $5$ & $6$ & $7$& $8$& $9$\\\hline
$l_4$ & $0$ & $-1$ & $1$& $0$ & $0$ & $0$ & $0$ & $0$ & $1$ & $-1$
\end{tabular}
\end{equation}
  The classic A-brane in the mirror Calabi-Yau manifold $X^{\ast}$ of X
determined by the additional charge vectors $(0,-1,1,0,0,0,0)$ is a
special Lagrangian submanifold of $X^{\ast}$ defined as
\cite{Aganagic:2000gs,Aganagic:2001nx,Kachru:2000an,Kachru:2000ih,Alim:2009rf}
\bqa -|x_1|^2+|x_2|^2=\eta\eqa where $x_i$ are coordinates on
$X^{\ast}$, $\eta$ is a K$\ddot{a}$hler moduli parameter with
$\hat{z}=\epsilon e^{-\eta}$ for a phase $\epsilon$.

  The differential operators, according to \eqref{operator}, can be
represented as

  \bqa
  \begin{split}
  &\mathcal{L}_1=\theta_1(\theta_1-2\theta_3)-12z_1(6\theta_1+5)(6\theta_1+1)\\
  &\mathcal{L}_2=(\theta_2-\theta_4)(\theta_2+\theta_4)+z_2(2\theta_2-\theta_3+1)(2\theta_2-\theta_3)\\
  &\mathcal{L}_3=\theta_3(-2\theta_2+\theta_3)+z_3(2\theta_3-\theta_1)(2\theta_3-\theta_1+1)\\
  &\mathcal{L}_4=\theta_4(\theta_2+\theta_4)+z_4\theta_4(\theta_4-\theta_2)
  \end{split}
  \eqa
  where $\theta_1=z_i\frac{d}{dz_i},~i=1,..,3$ and
  $\theta_4=\hat{z}\frac{d}{d\hat{z}}$.
   From the argument in sec.2, one obtains \bqa
   \begin{split}
 & \mathcal{L}_1=\mathcal{L}_1^{b}\Rightarrow\mathcal{L}_1^{b}=\theta_1(\theta_1-2\theta_3)-12z_1(6\theta_1+5)(6\theta_1+1)
 \\
  &\mathcal{L}_2=\mathcal{L}_2^{b}-\theta_2\theta_4^2\Rightarrow
  \mathcal{L}_2^{b}=\theta_2^2+z_2(2\theta_2-\theta_3+1)(2\theta_2-\theta_3)\\
  &\mathcal{L}_3=\mathcal{L}_3^{b}\Rightarrow\mathcal{L}_3^{b}=\theta_3(-2\theta_2+\theta_3)+z_3(2\theta_3-\theta_1)(2\theta_3-\theta_1+1).
  \end{split}
  \eqa
   Hence the inhomogeneous
  terms, by acting with $\theta_2\theta_4$ on the \eqref{solution},
  are
\bqa
  \begin{split}
  &\mathcal{L}_2^{b}W_1^{\pm}=\pm\frac{c}{2\pi^{\frac{5}{2}}}\sqrt{z_1z_2}z_3  \\
\end{split}
  \eqa

   For calculation of instanton corrections, one need to know mirror map. The closed-string periods are
\cite{Hosono:1993qy}
  \begin{equation} \label{Wp}
  \Pi(z) = \left(\begin{array}{c}
                    \omega_0(z,\rho)|_{\rho=0}\\
                    D_i^{(1)} \omega_0(z,\rho)|_{\rho=0}\\
                    D_i^{(2)} \omega_0(z,\rho)|_{\rho=0}\\
                    D^{(3)}   \omega_0(z,\rho)|_{\rho=0}\\
                  \end{array}
            \right).
\end{equation}
where $i=1,...,h_{21}(X^\ast)$,
  \bqa w_0=\sum c(n_i+\rho_i)z^{n_i+\rho_i} \qquad i=1,2,3
 \eqa
 \bqa
c(n_i+\rho_i)=\frac{\Gamma(\Sigma_{k=1}^{3}l_0^k(n_k+\rho_k)+1)}{\Pi_{i=1}^a\Gamma(\Sigma_{k=1}^3l_i^k(n_k+\rho_k)+1)}
 \eqa
 and \begin{equation}
  D_i^{(1)} := \partial_{\rho_i}, ~ D_i^{(2)} := \frac{1}{2} \kappa_{ijk} \partial_{\rho_j} \partial_{\rho_k},
  ~ D^{(3)} := -\frac{1}{6} \kappa_{ijk} \partial_{\rho_i} \partial_{\rho_j}
  \partial_{\rho_k}
\end{equation}
 $\kappa_{ijk}$ is intersection number of X.

  The flat coordinates in
 A-model at large radius regime are related to the flat coordinates of B-model at large complex structure regime by mirror map $t_i=\frac{\omega_i}{\omega_0},~\omega_i:= D_i^{(1)}
 \omega_0(z,\rho)|_{\rho=0}$
 \bqa
 \begin{split}
 &2\pi i t_1=\text{log}(z_1)+312z_1+58932z_1^2-(1-120z_1)z_3
 -\frac{3}{2}z_3^2+\mathcal{O}(z^3)\\
 &2\pi i t_2=\text{log}(z_2)+2z_2+3z_2^2+\mathcal{O}(z^3)\\
 &2\pi i
 t_3=\text{log}(z_3)+2z_3+3z_3^2+(120-240z_3)z_1+34380)z_1^2
 -\frac{3}{2}z_2^2-z_2+\mathcal{O}(z^3)
 \end{split}
 \eqa

   \begin{table}[!h]
\def\temptablewidth{1.0\textwidth}
\begin{center}
\begin{tabular*}{\temptablewidth}{@{\extracolsep{\fill}}c|ccccc}
$d_3=1$&&&&&\\
$d_1\backslash d_2$ & 1     &3     &5     &7    &9 \\\hline
1               & 8          & 0    & 0     &0   & 0     \\
  3              & -6784     & 0  & 0 & 0  & 0    \\
   5              & -2167824 &-82080  &82080& -82080  &82080    \\
   7              &-40065280 &66963200 &-66963200&66963200 &-66963200    \\
   9              & -9901094392&25006400960&-25006400960 &25006400960&-25006400960
       \end{tabular*}
       {\rule{\temptablewidth}{1pt}}
\end{center}
       \end{table}

\begin{table}[!h]
\def\temptablewidth{1.0\textwidth}
\begin{center}
\begin{tabular*}{\temptablewidth}{@{\extracolsep{\fill}}c|ccccc}
$d_3=2$&&&&&\\
 $d_1\backslash d_2$& 1     &3      &5     &7
&9\\\hline
1               & 3         & -3     & 0   &0   &0    \\
  3             & -2104     & -2104   & -4800  &4800&-4800      \\
   5            & 556278   & 1364046 &1933440 &-7477440  &-62400       \\
   7            & -1119704976&-649533072 &3072437120&1611134080   &4782650240    \\
   9            & -563398004013&1412362672931&646095703280 &879759458320 &1211151714800
       \end{tabular*}
       {\rule{\temptablewidth}{1pt}}
       \tabcolsep 0pt \caption{Ooguri-Vafa invariants $n_{(d_1,d_2,d_3)}$ for the on-shell
superpotential $W_1^+$ on the 3-fold $\mathbb{P}_{1,1,2,8,12}[24]$,
The horizonal coordinates represent $d_2$ and vertical coordinates
represent $d_1$.} \vspace*{-12pt}
\end{center}
       \end{table}

and we can obtain the inverse mirror map in terms of $q_i=e^{2\pi i
t_i}$
 \bqa
\begin{split}
&z_1=q_1-312q_1^2+87084q_1^3+q_1q_3-864q_1^2q_3+q_1q_2q_3+\mathcal{O}(q^4)\\
 &z_2=q_2-2q_2^2+3q_2^3+\mathcal{O}(q^4)\\
 &z_3=q_3-2q_3^2+3q_3^3-120q_1q_3+10260q_1^2q_3+q_2q_3-120q_1q_2q_3+600q_1q_3^2-4q_2q_3^2+\mathcal{O}(q^4).
\end{split}
 \eqa

 Using the modified multi-cover
formula\cite{Walcher:2006rs,Walcher:2009uj} for this case
  \bqa
\frac{W^{\pm}(z(q))}{w_0(z(q))}=\frac{1}{(2 i\pi)^2}\sum_{k~
 odd}~\sum_{d_3 , d_{1,2} odd\geq
0}n_{d_1,d_2,d_3}^{\pm}\frac{q_1^{kd_1/2}q_2^{kd_2/2}q_3^{kd_3}}{k^2},
\eqa the superpotentials $W^\pm$, at the critical points
$\hat{z}=1$, give Ooguri-Vafa invariants $n_{d_1,d_2,d_3}$ for the
normalization constants
  $c=1$, which are listed in Table 1. The three integers
  $(d_1,d_2,d_3)$ denote homology class.

\section{Superpotential of Hypersurface $X_{12}(1,1,1,3,6)$}
~~~~The $X_{12}(1,1,1,3,6)$ is defined as a degree 12 hypersurface
 in the ambient toric variety $W=P_{\Sigma(\triangle)}$ withe the
 vertices of the polyhedron $\triangle$
 \bqa
 \begin{split}
&\nu_1=(1,1,1,1),~~\nu_2=(-11,1,1,1),~~\nu_3=(1,-11,1,1),~~\nu_4=(1,1,-3,1)\\
&\nu_5=(1,1,1,-1)
\end{split}
\eqa
 The vertices of
the dual vertices $\triangle^{\ast}$ are \bqa \begin{split}
&\nu_1^{\ast}=(1,1,3,6),~~\nu_2^{\ast}=(-1,0,0,0),~~\nu_3^{\ast}=(0,-1,0,0),~~\nu_4^{\ast}=(0,0,-1,0)\\
&\nu_5^{\ast}=(0,0,0,-1),~~\nu_6^{\ast}=(2,1,0,0).
\end{split}
\eqa When considering the four-fold on which the dual F-theory
compactification, the extend vertices of the enhanced polyhedron
$\underline{\triangle^{\ast}}\supset\triangle^{\ast}$ can be
constructed according to \eqref{four} with two extra points as
follows \bqa
\underline{\nu_i^{\ast}}=(\nu_i^{\ast};0),~\underline{\nu_7^{\ast}}=(\nu_1^{\ast};1),~\underline{\nu_8^{\ast}}=(\nu_2^{\ast};1),~i=1,...,6.
\eqa This threefold has three complex parameters, but only two
moduli can be represented as monomial deformation
\cite{Hosono:1993qy}.

     The toric hypersurface is defined as zero locus of $P$:  \bqa
\begin{split}
P=x_1^{12}+x_2^{12}+x_3^{12}+x_4^4+x_5^{2}+\psi x_1x_2x_3x_4x_5+\phi
x_1^4x_2^4x_3^4
\end{split}
\eqa where $\psi=z_1^{-\frac{1}{4}}z_2^{-\frac{1}{12}}$,
$\phi=z_2^{-\frac{1}{3}}$. The GLSM charge vectors in this case are
\cite{Hosono:1993qy}
\begin{equation}
\begin{tabular}{c|c c c c c c c}
~  & $0$ & $1$& $2 $ & $3$ & $4$ & $5$ & $6$ \\\hline
$l_1$ & $-4$ & $0$ & $0$ & $0$ & $1$ & $2$ & $1$ \\
$l_2$ & $0$ & $1$ & $1$& $1$ & $0$ & $0$ & $-3$
\end{tabular}
\end{equation} On the mirror manifolds, the
Greene-Plesser orbifold group acts as $x_i\rightarrow
\lambda_k^{g_{k,i}}x_i$ with weights \bqa
\mathbb{Z}_6:~g_1=(1,-1,0,0,0),~~~\mathbb{Z}_4:~g_2=(0,1,2,1,0)\eqa
where we denotes $\lambda_{1}^6=1,~\lambda_{2}^4=1$.
   We consider those curves
  \bqa
  \begin{split}
  C_{\alpha,\pm}=\{x_2=&\xi_1
  x_1,x_5=\xi_2x_3^2-\frac{\psi}{2}x_1x_2x_3x_4,x_4=\psi^2\alpha(x_1x_2x_3x_4)^2\}\\
  &\xi_1^{12}=\xi_2^2=-1,~~\alpha^3-\frac{1}{4}\alpha^2+\frac{\phi}{\psi^6}=0.
  \end{split}
  \eqa
  here with the identifications for different choice of
  $(\xi_1,\xi_2,\alpha)$ but for $\xi_1^6=\pm i$ for fixed
  $(\xi_2,\alpha)$.

    The family of divisor we calculated is \bqa
Q(\mathcal{D})=x_2^{12}+\hat{z}x_1^{12}\eqa where the critical
points is at $\hat{z}=1$. Solving the intersection
$P=0=Q(\mathcal{D})$ one obtains
  \bqa
  P_{\mathcal{D}}=(x_2')^{12}+x_3^{12}+x_4^{4}+x_5^2+\psi'x_2'x_3x_4x_5+\phi' (x_2')^4x_3^4
  \eqa
where $x_2'=x_2^2$, $\psi'=u_1^{-\frac{1}{4}}u_2^{-\frac{1}{12}}$
and $\phi'=u_2^{-\frac{1}{3}}$ are expressed in terms of new
parameters as \bqa u_1=z_1 \qquad
u_2=-\frac{z_2}{\hat{z}}(1-\hat{z})^2. \eqa The family of divisor
$\mathcal{D}$ as a $K3$ surface associated with GLSM charge vectors
\begin{equation}
\begin{tabular}{c|c c c c c c}
~  & $0$ & $2 $ & $3$ & $4$ & $5$ & $6$ \\\hline
$\hat{l}_1$ & $-4$ & $0$ & $0$ & $1$ & $2$ & $1$ \\
$\hat{l}_2$ & $0$ & $2$ & $1$ & $0$ & $0$ & $-3$
\end{tabular}
\end{equation}
has two algebraic moduli.

    Besides the regular solutions the period on this K3 surface has two extra
    forms
  \bqa
  \begin{split}
  \eqlabel{solution}
  &\pi_1(u_1,u_2)=\frac{c_1}{2}B_{\{\hat{l_1},\hat{l_2}\}}(u_1,u_2;0,\frac{1}{2})\\
  &\pi_2(u_1,u_2)=\frac{c_2}{2}B_{\{\hat{l_1},\hat{l_2}\}}(u_1,u_2;\frac{1}{2},\frac{1}{2})
  \end{split}
  \eqa
  where $c_{1,2}$ are some normalization constants not determined by the
  differential operator.

 The additional GLSM charge vectors corresponding to the divisor
   are
       \begin{equation}
\begin{tabular}{c|c c c c c c c c c c}
 & $0$ & $1$& $2$& $3 $ & $4$ & $5$ & $6$   &$7$ &$8$\\\hline
$l_3$ & $0$ & $-1$ & $1$& $0$ & $0$ & $0$  & $0$ & $1$ & $-1$
\end{tabular}
\end{equation}
 The classic A-brane in the mirror Calabi-Yau manifold $X^{\ast}$ of X
determined by the additional charge vectors $(0,-1,1,0,0,0,0)$ is a
special Lagrangian submanifold of $X^{\ast}$ defined as
\cite{Aganagic:2000gs,Aganagic:2001nx,Kachru:2000an,Kachru:2000ih,Alim:2009rf}
\bqa -|x_1|^2+|x_2|^2=\eta\eqa where $x_i$ are coordinates on
$X^{\ast}$, $\eta$ is a K$\ddot{a}$hler moduli parameter with
$\hat{z}=\epsilon e^{-\eta}$ for a phase $\epsilon$.

  The differential operators, according to \eqref{operator}, can be
obtained as
  \bqa
  \begin{split}
  &\mathcal{L}_1=\theta_1(\theta_1-3\theta_2)-4z_1(4\theta_1+3)(4\theta_1+1)\\
  &\mathcal{L}_2=\theta_2(\theta_2-\theta_3)(\theta_2+\theta_3)+z_2(3\theta_2-\theta_1)(3\theta_2-\theta_1+1)(3\theta_2-\theta_1+2)\\
  &\mathcal{L}_3=\theta_3(\theta_2+\theta_3)+z_3\theta_3(\theta_2-\theta_3)
  \end{split}
  \eqa
   where $\theta_1=z_i\frac{d}{dz_i},~i=1,..,3$ and
  $\theta_4=\hat{z}\frac{d}{d\hat{z}}$.

    Furthermore one has \bqa
    \begin{split}
 &\mathcal{L}_1=\mathcal{L}_1^{b}\Rightarrow\mathcal{L}_1^{b}=\theta_1(\theta_1-3\theta_2)-4z_1(4\theta_1+3)(4\theta_1+1)\\
  &\mathcal{L}_2=\mathcal{L}_2^{b}-\theta_2\theta_4^2\Rightarrow
  \mathcal{L}_2^{b}=\theta_2^2+z_2(3\theta_2-\theta_1)(3\theta_2-\theta_1+1)(3\theta_2-\theta_1+2).
  \end{split}
  \eqa

 Following the section 3, The on-shell superpotentials
 $W^{\pm}$ for this manifold can be obtained as
 \bqa
 W_a^{\pm}(z_1,z_2)=\mathcal{W}_a^{\pm}\mid_{\hat{z}=1}.
 \eqa
 The explicit form of on-shell superpotentials are
 \bqa
 \begin{split}
&W_1^{\pm}=\mp\frac{c_1}{8}\prod_{n_1=0,n_2=0}\frac{\Gamma(1+4n_1)z_1^{n_1}z_2^{n_2+\frac{1}{2}}}{\Gamma^3(n_2+\frac{3}{2})\Gamma(n_1+1)\Gamma(2n_1+1)\Gamma(n_1-3n_2-\frac{1}{2})}\\
&W_2^{\pm}=\mp\frac{c_2}{8}\prod_{n_1=0,n_2=0}\frac{\Gamma(3+4n_1)z_1^{n_1+\frac{1}{2}}z_2^{n_2+\frac{1}{2}}}{\Gamma^3(n_2+\frac{3}{2})\Gamma(n_1+\frac{3}{2})\Gamma(2n_1+2)\Gamma(n_1-3n_2)}.
\end{split}
\eqa

 The flat coordinates $t_i$ in A-model sides for this case are
 \bqa
 \begin{split}
 &2\pi i t_1=\text{log}(z_1)+40z_1+1076z_1^2+(2-36z_1)z_2-15z_2^2+\mathcal{O}(z^3)\\
 &2\pi i
 t_2=\text{log}(z_2)-6z_2+45z_2^2+36z_1+1458z_1^2+108z_1z_2\mathcal{O}(z^4),
 \end{split}
 \eqa
 and the inverse mirror map in terms of $q_i=e^{2\pi i t_i(z)}$ are
 \bqa
\begin{split}
&z_1=q_1-40q_1^2+1324q_1^3-2q_1q_2+268q_1^2q_2+5q_1q_2^2+\mathcal{O}(q^4)\\
 &z_2=q_2+6q_2^2+9q_2^3-36q_1q_2-468q_1q_2^2+630q_1^2q_2+\mathcal{O}(q^4).
 \end{split}
 \eqa
  Using the modified multi-cover
formula\cite{Walcher:2006rs,Walcher:2009uj} for this case
 \bqa
\frac{W^{\pm}(z(q))}{w_0(z(q))}=\frac{1}{(2\pi i)^2}\sum_{k~
 odd}~\sum_{d_2~odd,~d_{1}\geq
0}n_{d_1,d_2}^{\pm}\frac{q_1^{kd_1}q_2^{kd_2/2}}{k^2}, \eqa
 the on-shell superpotentials $W^\pm$, at the critical point $\hat{z}=1$, give
 Ooguri-Vafa invariants $n_{d_1,d_2}$ for the normalization constants
  $c_1=c_2=1$, which are listed in Table. 2 and 3.

\begin{table}[!h]
\def\temptablewidth{1.0\textwidth}
\begin{center}
\begin{tabular*}{\temptablewidth}{@{\extracolsep{\fill}}c|ccccc}
$d_1\backslash d_2$ & 1     &3     &5     &7  &9  \\\hline
0               & 1    &-1    & 5    &-42  &429  \\
1              & -54  & 54  & -486   & 5454&-116316  \\
2              & -1107  & 0 &17199 &  -293463 &7513614\\
3             &-10686  & 20088&8520336  & -14892288 & 65638798 \\
4         &-71496&43155864&-1157860257&12330791559&-150518794344
       \end{tabular*}
       {\rule{\temptablewidth}{1pt}}
\tabcolsep 0pt \caption{Ooguri-Vafa invariants
$\frac{1}{2}n_{(d_1,d_2)}$ for the on-shell superpotential $W_1^{+}$
on the three-fold $\mathbb{P}_{1,1,1,3,6}[12]$. The horizonal
coordinates represent $d_2$ and vertical coordinates represent
$d_1$.} \vspace*{-12pt}
\end{center}
       \end{table}

\begin{table}[!h]
\def\temptablewidth{1.0\textwidth}
\begin{center}
\begin{tabular*}{\temptablewidth}{@{\extracolsep{\fill}}c|ccccc}
$d_1\backslash d_2$ & 1     &3     &5     &7    &9  \\\hline
1               & 0       & 0     & 0     & 0 & 0   \\
  3              & -320    & 0     & 0 & 0 & 0  \\
   5              & -3456   & 6912 & -17280 & 110592 &13768704  \\
   7              &-29376    & -255744&1081728 &-10596096&287378496   \\
   9              &-166528  & 599128640&8281234560&865215488&-258928821888
       \end{tabular*}
       {\rule{\temptablewidth}{1pt}}
       \tabcolsep 0pt \caption{Ooguri-Vafa invariants $\frac{1}{2}n_{(d_1,d_2)}$ for the on-shell
superpotential $W_2^{+}$ on the three-fold
$\mathbb{P}_{1,1,1,3,6}[12]$. The horizonal coordinates represent
$d_2$ and vertical coordinates represent $d_1$.} \vspace*{-12pt}
\end{center}
       \end{table}

\section{Summary}
  ~~~~In this paper, we constructed the generalized hypergeometric GKZ systems for two Calabi-Yau manifolds with three parameters and D-brane wrapped on a divisor with single open-string moduli. Furthermore, we calculate the D-brane
superpotentials which give rise to the flux superpotential
$\mathcal{W}_{GVW}$ of the dual F-theory compactify on the relevant
Calabi-Yau fourfold in the limit of $g_s\rightarrow0$. The
superpotentials are depend on bulk and open deformation moduli. By
mirror symmetry, we also compute the Ooguri-Vafa invariants from
A-model expansion.

  The generalized hypergeometric GKZ system are closely related to the variation of mixed Hodge structure on relative cohomology group $H^3(X,Z)$. The cohomology group
  $H^3(X,Z)$ can be viewed as the fiber of vector bundle over the
  deformation space $\mathcal{M}$. Similar to closed-string, there is a flatness and integrability of
the Gauss-Manin connection which provided a powerful approach to
study the geometry of B-model and compute the superpotentials. The
connection in flat coordinates displays, in fact, an quantum ring
structure and predictions of corrections of the disc instantons.

  For the dual F-theory superpotential \cite{Alim:2010za,Jockers:2009ti,Berglund:2005dm,Grimm:2009ef}, which also are solutions of the generalized GKZ
  system, is not only related to the D-brane superpotential for Type II compactification, but
  also to superpotential for
  heterotic theory compactification \cite{Berglund:1998ej,Grimm:2009sy}. In type II/F-theory compactification, the vacuum structure is determined by
  the superpotentials, whose second derivative gives the chiral ring structure. The quantum cohomology ring structure comes from the world-sheet instanton corrections and
  space-time instanton corrections\cite{Lerche:2002ck,Lerche:2002yw}. In fact, the more general vacuum
  structure of type II/F-theory/heterotic theory compactification can be tackled in Hodge variance approach.

  We will study the extremal transition and monodromy problems as
  well as
  D-brane in general case. We also try to calculate the D-brane superpotential
  with the method of
  $A_{\infty}$ structure of the derived category $D_{\text{coh}}(X)$ and path algebras
  of quivers.

\section*{Acknowledgments}
~~~~The author (Yang) would like to appreciate Profs. Bo-Yuan Hou,
Ke-Feng Liu, Zhong-Qi Ma, Shi-Kun Wang and Ke Wu for many helps,
also appreciate Prof. Hans Jockers for valuable correspondence. The
work is supported by the NSFC (11075204) and President Fund of GUCAS
(Y05101CY00).

\end{document}